\documentclass[12pt,preprint]{aastex}

\newcommand{\xte}{{\it RXTE}}

\newcommand{\saxsource}{SAX~J1748.9$-$2021}

\def\lapp{\ifmmode\stackrel{<}{_{\sim}}\else$\stackrel{<}{_{\sim}}$\fi}
\def\gapp{\ifmmode\stackrel{>}{_{\sim}}\else$\stackrel{>}{_{\sim}}$\fi}

\def\nuddd{\ifmmode\stackrel{\bf \,...}{\textstyle \nu}\else$\stackrel{\,...}{\textstyle \nu}$\fi}
\def\nudddd{\ifmmode\stackrel{\bf \,....}{\textstyle \nu}\else$\stackrel{\,....}{\textstyle \nu}$\fi}

\begin{document}
\title{Discovery of 442-Hz Pulsations from an X-ray  Source in the Globular
Cluster NGC~6440}
\author{Fotis P. Gavriil\altaffilmark{1}, Tod E. Strohmayer, Jean H. Swank, Craig B. Markwardt\altaffilmark{2,3}}
\altaffiltext{1}{NPP Fellow; Oak Ridge Associated Universities, Oak Ridge, TN}
\altaffiltext{2}{CRESST}
\altaffiltext{3}{Department of Astronomy, University of Maryland, College Park, MD 20742.}
\affil{NASA Goddard Space Flight Center, Astrophysics Science Division, Code 662, X-ray Astrophysics Laboratory, Greenbelt, MD 20771, USA} 

\begin{abstract}
We report on the serendipitous discovery of a 442-Hz pulsar during a
\textit{Rossi X-ray Timing Explorer} (\xte) observation of the
globular cluster NGC~6440.  The oscillation is detected following a
burst-like event which was decaying at the beginning of the
observation. The time scale of the decay suggests we may have seen the
tail-end of a long-duration burst.  Low-mass X-ray binaries (LMXBs)
are known to emit thermonuclear X-ray bursts that are sometimes
modulated by the spin frequency of the star, the so called burst
oscillations.  The pulsations reported here are peculiar if
interpreted as canonical burst oscillations.  In particular, the pulse
train lasted for $\sim$500 s, much longer than in standard burst
oscillations. The signal was highly coherent and drifted down by
$\sim$2$\times10^{-3}$Hz, much smaller than the $\sim$Hz drifts
typically observed during normal bursts.  The pulsations are
reminiscent of those observed during the much more energetic
``superbursts'', however, the temporal profile and the energetics of
the burst suggest that it was not the tail end nor the precursor
feature of a superburst.  It is possible that we caught the tail end
of an outburst from a new `intermittent'' accreting X-ray millisecond
pulsar, a phenomenon which until now has only been seen in
HETE~J1900.1$-$2455 \citep{gmk+07}.  We note that \citet{kzht03}
reported the discovery of a 409.7~Hz burst oscillation from
\saxsource, also located in NGC~6440.  However, \textit{Chandra X-ray
  Observatory} imaging indicates it contains several point-like X-ray
sources, thus the 442~Hz object is likely a different source.
\end{abstract}

\section{Introduction}
\label{sec:intro}
The discovery of millisecond spin periods of neutron stars in Low Mass
X-ray Binaries (LMXBs) with the \textit{Rossi X-ray Timing Explorer}
(\xte) has helped elucidate the nature of these sources.
Neutron star LMXBs consist of a neutron star accreting from a low mass
companion. As material (mostly H and He) is accreted onto the star and
gets compressed, it eventually ignites and burns unstably
\citep[see][]{sb07}. This phenomenon is observed as a Type I X-ray
Burst. Type I X-ray bursts have been observed from over $\sim$70 LMXBs
\citep[see][and references therein]{lph06}. The recurrence time of these
bursts varies but in some cases it can be as frequent as every few
hours \citep[see][for examples]{gm07}. Occasionally it is possible to
observe the spin of the neutron star modulating the burst emission --
the so called ``burst oscillations'' \citep{szs+96}. Burst
oscillations have been observed from $\sim$18 LMXBs \citep[see][for
  examples]{gm07}.

X-ray bursts from LMXBs have been discovered which are $\sim$1000
times longer, and thus, that much more energetic, than canonical Type
I X-ray bursts. They are aptly named ``superbursts''. Superbursts are
believed to occur by the unstable burning of carbon
\citep{sb02,cb01}. \citet{sm02} discovered highly coherent pulsations
during a superburst from 4U~1636$-$53. The pulse train lasted for
$\sim$800~s, as opposed to the $\sim$10~s long pulse trains observed
in Type I X-ray bursts. Superbursts are much rarer than Type I X-ray
bursts. Thus far ten have been observed from eight LMXBs
\citep[see][]{zcc04}.


\section{Observations}
\label{sec:observations}

\begin{figure}
\plotone{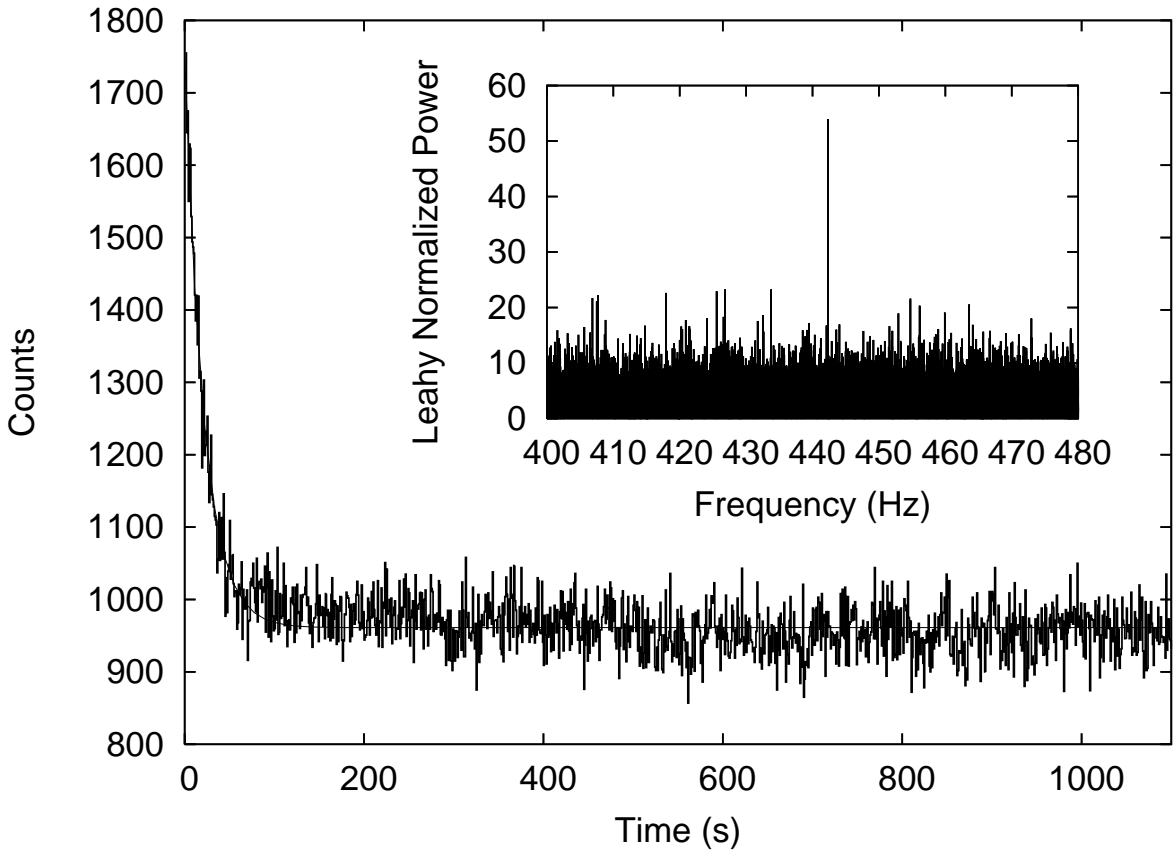}
\caption{Burst lightcurve binned with 1-s resolution and using the
  full PCA bandpass. Inset: ``Leahy'' normalized \citep{lde+83} power
  spectrum of the entire observation. Notice the significant peak at
  442 Hz. \label{fig:lc}}
\end{figure}

Between 2005 March 7 and 2005 July 21 the PCA Galactic Bulge Scan
Survey\footnote{\texttt{http://lheawww.gsfc.nasa.gov/users/craigm/galscan/}}
discovered an outburst from the direction of the Globular Cluster
NGC~6440. A follow up 1.8-ks long pointed \xte\ observation
(observation identification number 91050-03-07-00) was initiated on
2005 June 14.  The data presented here were acquired from the
Proportional Counter Array (PCA) on board \xte.  The PCA consists of
five identical and independent proportional counter units (PCUs). Each
PCU is a 90\% Xenon/10\% Methane gas filled proportional counter.
Each PCU has a collimated 1$\degr$$\times$1$\degr$ field-of-view, 256
spectral channels in the 2--60~keV range, and a limiting temporal
resolution of $\sim$1~$\mu$s.  Only two PCUs (PCU~0 and PCU~2) were
operational throughout the span of the observation.  The data were
taken in \texttt{E\_125us\_64M\_0\_1s} mode, which returns events to a
limiting resolution of 125~$\mu$s and with moderate (64 channels as
opposed to the full 256 channels) spectral resolution. This mode was
used because it is not as susceptible to buffer overflows during high
countrate data as compared to the less restrictive modes
(e.g. \texttt{GoodXenon}).  Using the pointing of the observation
(17$^{\mathrm{h}}$48$^{\mathrm{m}}$52\fs8, -20\degr 21\arcmin
32\arcsec) and the planetary ephemeris DE200, the times of raw events
were corrected to the solar system barycenter.  Binning the events
into a 1-s resolution lightcurve reveals a quickly decaying burst
(Fig.~\ref{fig:lc}).

\section{Analysis and Results}
\label{sec:results}


This burst was reminiscent of canonical Type~I X-ray bursts from
LMXBs, therefore we searched this event for burst oscillations.  We
rebinned the raw events using the full spectral band pass into a time
series with 0.5/1024~s resolution, which yields an equivalent Nyquist
frequency of 1024~Hz. A Leahy normalized \citep[see][]{lde+83} power
density spectrum (PDS) is displayed in Figure~\ref{fig:lc} (inset). A
prominent peak is clearly seen at 442~Hz. The probability of this peak
occurring by chance after accounting for the number of trials (the
total number of frequency bins in our PDS) is $\sim$$2\times 10^{-9}$.

\subsection{The NGC~6440 Field} 
\label{sec:FOV}

NGC~6440 harbors the bright X-ray transient \saxsource, which
exhibited a 409.7~Hz burst oscillation \citep{kzht03}. However, there
are many other X-ray sources in the cluster (see Fig~\ref{fig:chandra
  image}). \textit{Chandra X-ray Observatory} (\textit{CXO}) imaging by \citet{plv+02} revealed 24 X-ray
sources in NGC~6440, and they concluded that 4--5 of these sources are
likely quiescent LMXBs. Thus, the phenomenon we discovered is most
likely from a different object in the cluster.

\begin{figure}
\includegraphics[width=\columnwidth]{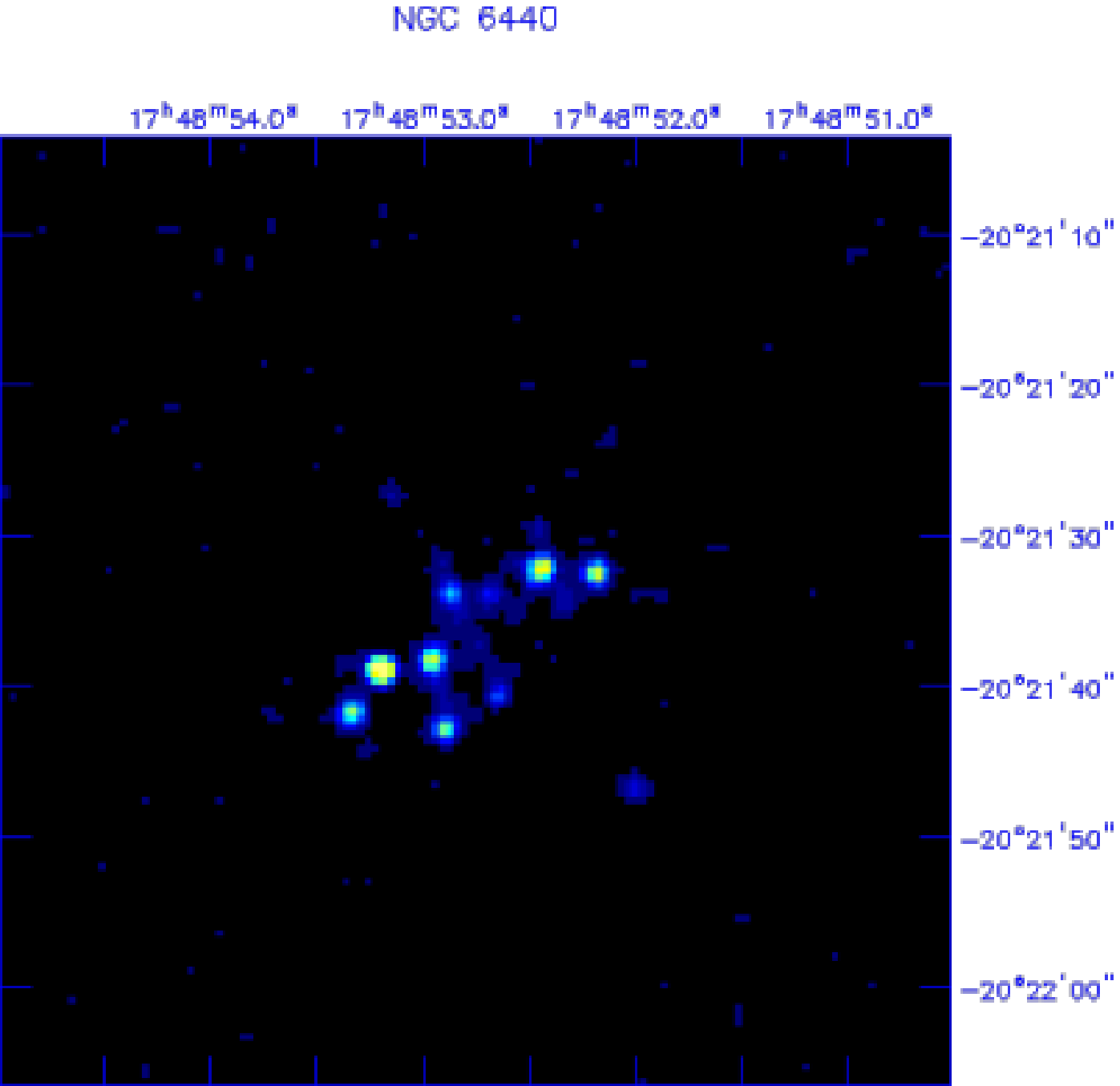}
\caption{Archival \textit{CXO} image of
  NGC~6440. Notice that the cluster contains many X-ray point sources.
\label{fig:chandra image}}
\end{figure}

\subsection{Timing Analysis}
\label{sec:timing}
\begin{figure*}
\centerline{\includegraphics[height=1.5\columnwidth, angle=270]{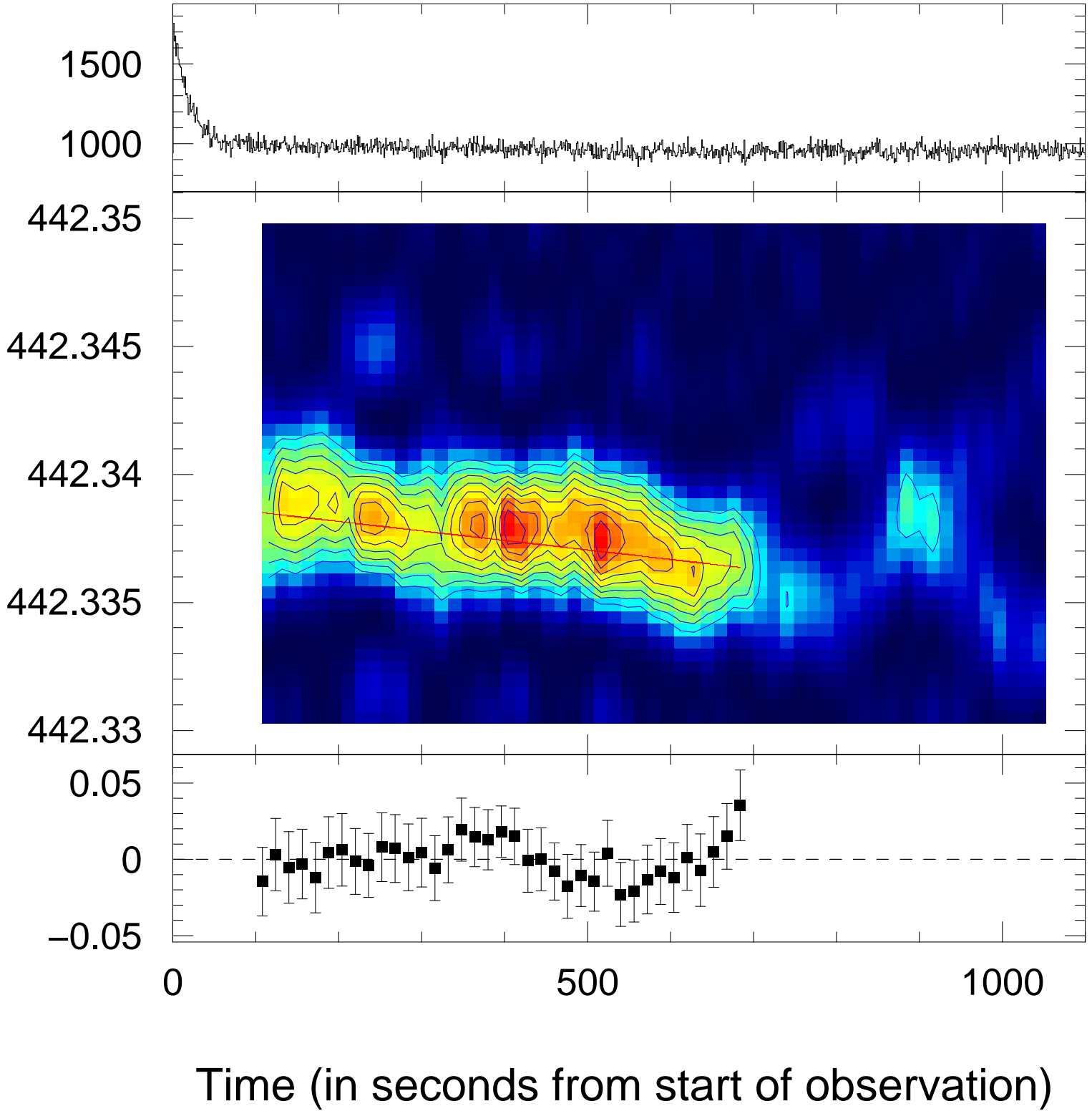}}
\figcaption{ Pulse Frequency Evolution. Top:
  The burst lightcurve as shown in Fig.~\ref{fig:lc}. Middle: Dynamic
  $Z^2$ power spectrum. Notice that the pulsations remain highly
  significant (with $Z_1^2>16$) for 576~s. The solid line represents our
  best-fit frequency model consisting of just a frequency and a
  frequency derivative term. Bottom: Phase residuals after subtracting
  our best fit frequency model. Notice that the pulsations are highly
  coherent.
\label{fig:frequency evolution}}
\end{figure*}

To study  the frequency evolution of the pulse we  calculated a
dynamic $Z^2$ statistic .The $Z^2$ statistic is analogous to the Fourier power
spectrum, with the advantage that the data need not be binned, thus
allowing us to oversample our data.
The $Z^2$ statistic is defined as:
\begin{equation}
Z^2_{N_{\mathrm{harm}}} = \frac{2}{N_{\gamma}}\sum_{k=0}^{N_{\mathrm{harm}}} \sum_{i=0}^{N_{\gamma}} \left| e^{i 2\pi k \nu_j t_i }\right|^2
\end{equation}
where $N_\gamma$ is the total number of photons in each interval,
$N_{\mathrm{harm}}$ is the total number of harmonics that one deems
significant, $\nu_j$ is the frequency searched over, and $t_i$ is the
event time. The factor in front of the summations normalizes the $Z^2$
statistics in an analogous way to Leahy normalizing a PDS.
We calculated the dynamic $Z^2$ power spectrum using a 200-s long
window, which was translated across the observation with a step size
of 16~s.  Our dynamic $Z^2$ power spectrum is displayed in
Fig.~\ref{fig:frequency evolution}.  Notice that the pulsations where
highly significant for 576~s. As is common for canonical X-ray bursts,
the frequency drifts; however, unlike the large $\sim$1-2~Hz drifts
seen in those, the pulsation here drifts down in frequency only by
$\sim$2.1$\times 10^{-3}$~Hz in 576~s.

\begin{figure}
\plotone{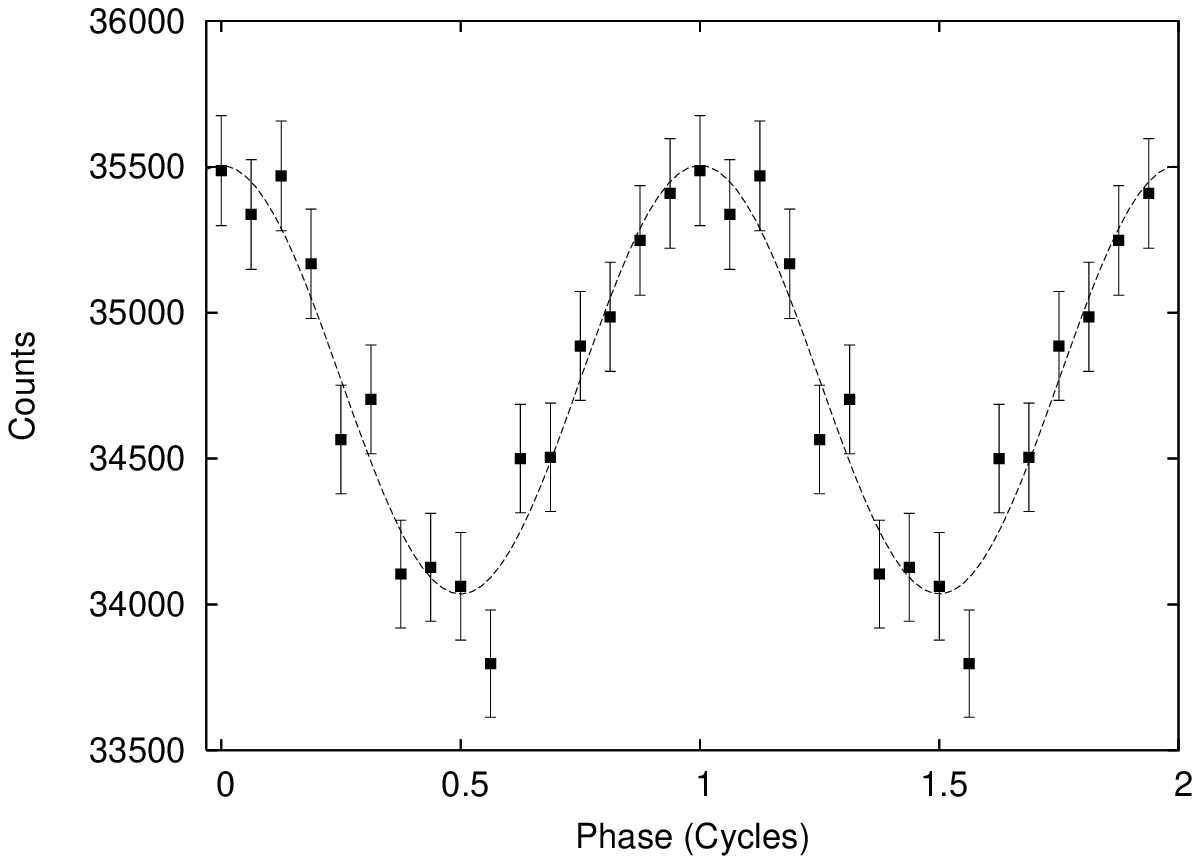} 
\figcaption{ The pulse profile obtained by
  folding all events in the 576-s long pulse train (i.e., the interval
  where the pulsations were highly significant as determined from the
  $Z^2$ power spectrum) on the ephemeris determined in our timing
  analysis. The pulse shape is highly sinusoidal. The solid line
  represents the best-fit sinusoid. The fit yielded $\chi^2$=0.76 for
  13 degrees of freedom, affirming the absence of any harmonic
  content.
\label{fig:profile}}
\end{figure}

To further quantify this frequency evolution we determined how the
phase of the pulsations varies as a function of time.  For each interval
with $Z_1^2>16$ we generated a pulse profile by folding the events in
that interval on the frequency determined from the $Z^2$ statistic. We
then cross correlated these pulse profiles with a sinusoid of fixed
phase to determine the pulse phase as a function of time. We can model
the pulse phase ($\phi$) at a given time ($t$) by the following Taylor
expansion:
\begin{equation}
\phi(t) = \phi(t_0) + \nu(t_0) (t - t_0) + \frac{1}{2} \dot{\nu}(t_0)
(t-t_0)^2 + \cdots,
\label{eq:phase}
\end{equation}
where $\nu$ is the barycentric frequency, $\dot{\nu}$ is the frequency
derivative, and $t_0$ is some reference epoch. We were able to
“whiten” our phase residuals with just a frequency derivative (see
Fig.~\ref{fig:frequency evolution}, bottom). We find
$\nu=442.33850(5)$~Hz, $\dot{\nu}=3.7(2)\times 10^{-6}$~Hz~s$^{-1}$ at
$t_0$=53535.463215~MJD~(UTC). The phase residuals after subtracting
the model given by Eq.~\ref{eq:phase} are shown in
Fig.~\ref{fig:frequency evolution}~(bottom panel).  Notice that the
pulse train lasts for 576~s; this is quite long when compared to those
of Type I X-ray bursts.  We folded all the events within the 576~s
pulse train using our best fit $\nu$ and $\dot{\nu}$, and the resulting
pulse profile is shown in Fig.~\ref{fig:profile}. Notice that the pulse
profile is highly sinusoidal. Fitting a sinusoid to the pulse profile
returned a reduced $\chi^2$ value of 0.76 for 13 degrees of freedom,
affirming the absence of any harmonic content. From the fit we obtain a fractional pulse amplitude of 2.1$\pm$0.1\%.

It is possible that the frequency drift we have observed
is entirely due to orbital motion. The observed spin frequency of a
neutron star in a circular orbit is given by
\begin{equation}
\nu = \nu_0 \left [ 1 - \frac{v_{\mathrm{ns}}\sin i}{c}\sin \left(
  \frac{2\pi t}{T} + \phi_0 \right) \right],
\label{eq:orbit}
\end{equation}
where, $\nu_0$ is the barycentric frequency at time $t=t_0$,
$v_{\mathrm{ns}}\sin i$ is the projected velocity of the neutron star,
$T$ is the orbital period, and $\phi_0$ is the orbital phase at time
$t=t_0$. Unfortunately we do not have a long enough data set to place
interesting constraints on $T$ or $v_{\mathrm{ns}}\sin i$; however, we
can determine whether orbital modulation is consistent with the
observed frequency drift.  For example, if we assume, for solely
demonstrative purposes, that this system has a projected velocity
comparable to the one found by \citet{sm02} for 4U~1636$-$53
($v_{\mathrm{ns}}\sin i=136$~km~$^{-1}$), then fitting Eq.~\ref{eq:orbit} to our frequency
time series yields a good fit with a reasonable orbital period, i.e.
$T=$7$\pm$1~hr. Hence, binary motion can account for the observed
frequency drift. Observing another outburst  would be
essential in ultimately constraining the orbital parameters of this
system.

\subsection{Spectral Analysis}
\label{sec:spectral}

The timing analysis seems to suggest that this event is in fact very
different from a canonical burst oscillation. To study it further we
performed a spectral analysis.  We analyzed the spectral evolution of
the burst by breaking the burst up into ten intervals. The exposure of
each interval was selected by demanding that each one contained an
equal number of photons. Using the last interval as an estimator of
the persistent emission, we fit each interval to a simple
photoelectrically absorbed blackbody model while holding the column
density fixed to the value found by \citet{plv+02} for the optical
reddening.  We find only subtle evidence for spectral softening,
significant only at the 1-$\sigma$ level.

Using the distance estimate to the cluster found by \citet{obb94},
$d=8.5\pm0.4$~kpc, we were able to calculate the luminosity of the
burst.  At the start of the observation the flux was
$\sim$2$\times$$10^{37}$~erg~s$^{-1}$, which is $\sim$0.1$L_{Edd}$,
where $L_{Edd}$ is the Eddington luminsosity for a neutron star.  The
burst lightcurve was well fit by an exponential with a decay timescale
of 22~s. Unfortunately the observation only caugth the tail end of the
burst-like event, we could not therefore characterize the peak
luminosity of the burst.

\section{Discussion}
\label{sec:discussion}

\subsection{Super Burst?}

The timing properties of the pulsations are reminiscent of those
observed during superbursts. In particular the highly coherent
pulsations and the long pulse trains.  The superburst interpretation
would be very exciting given the rarity of such events. However, we
could not unambiguously determine this is the case given that we only
caught the tail end of the burst. To study this phenomenon further we
extrapolated our luminosity model determined in our spectral analysis
before the start time of the observation. We find that the burst
reaches the Eddington luminosity $\sim$50~s before the start of the
observation.  Now, if we assume that we caught the tail of the
outburst it is very surprising that it dropped from $\sim$$L_{Edd}$
down to $\sim$.01$L_{Edd}$ in $\sim$77~s.  This is much too fast for a
superburst-like event. If, on the other hand, this is a precursor
events, such as those seen in superbursts, than it is unclear why the
persistent flux did not rise back to $\sim$$L_{Edd}$ shortly after as
observed in other superbursts. Thus, the spectral analysis of this
event suggests it is not a superburst.

\subsection{Intermittent Pulsar?}
The timing properties of the pulsations  share all the hallmarks of those observed during superbursts, but the
spectral analysis makes this interpretation unlikely.
If this event is not a superburst, then could it be that these highly
coherent pulsations have eminated from an accretion-powered
millisecond pulsar?  Recently \citet{gmk+07} discovered
``intermittent'' pulsations from the accretion-powered millisecond
pulsar (AMP) HETE~J1900.1$-$2455. They found that the properties of
this pulsar differed from those of the other six known AMPs. For
example, the pulsations were only present in the first few months of
its outburst as opposed to other AMPs which show pulsations throughout
\citep[see][for a review]{wij04}. In addition, the pulse amplitude
increased at three points in time that were almost coincident with the
times of thermonuclear bursts. The pulsed amplitude subsequently
decaded after the bursts.  Thus, it is plausible that we have observed
a similar phenomenon. Following the theoretical work of \citet{czb01}
and \citet{pm06}, \citet{gmk+07} suggest that the reason why the pulsations in
HETE~J1900.1$-$2455 sometimes ``switch off'' is because the magnetic
field is burried by accreted material, thus it cannot channel the flow
which gives rise to the pulsations in the first place

If the event we have discovered is another example of these
intermittent pulsations from an unidentified source in NGC~6440, then
a future outburst should reveal more long pulse trains, and if they
happen to be correlated with the times of Type-I X-ray bursts then
this would solidify the connection between this source and
HETE~J1900.1$-$2455.


\section{Conclusions}
We discovered a 442~Hz pulsation during a burst from an X-ray source
in NGC~6440.  We could not establish the energetics of the burst as
\textit{RXTE} only caught the tail end of the event, but based on the
timing properties of the pulsation we conclude that it was not a
canonical burst oscillation. In particular we find that: The pulse
train lasted for 576~s, as opposed to the $\sim$10~s long pulse trains
observed in Type I X-ray bursts.  The signal was highly coherent and
drifted down only by $2.1\times10^{-3}$~Hz, as opposed to Type-I X-ray
burst which exhibit drifts which are three orders of magnitude
larger. The pulsations share all the properties of those observed
during superbursts; however, the complete energetics of this event
make this interpretation implausible. We conclude that the long pulse
train is most likely an ``intermittent'' pulsation from an accreting
millisecond pulsar such as those seen only in HETE~J1900.1-2455 thus
far \citep{gmk+07}. We note that NGC~6440 is known to harbor a
transient X-ray source with a 409.7~Hz spin frequency \citep{kzht03}
However, X-ray imaging of the cluster by \citet{plv+02} has shown that
NGC~6440 contains many X-ray sources. Thus, the burst presented here
most likely emanated from a different object in the cluster.

\acknowledgements
FPG is supported by the NASA Postdoctoral Program
administered by Oak Ridge Associated Universities at NASA Goddard
Space Flight Center. This research has made use of data obtained
through the High Energy Astrophysics Science Archive Research Center
Online Service, provided by the NASA/Goddard Space Flight Center.

\end{document}